\documentclass{jaa}
\usepackage{natbib}
\usepackage{graphicx}
\usepackage{multirow}
\usepackage{subfigure}
\usepackage{xcolor}
\usepackage[colorlinks=true,linkcolor=blue,citecolor=blue]{hyperref}
\begin{document}
%\sloppy

%%paper title
%%For line breaks \\ can be used within title
\title{A Polarization Study of 3 Blazars using the uGMRT at $\sim$600~MHz}

%%author names are separated by comma (,)
%%use \and before the last author name
%%use a * along with the number separated by comma
%% for the  author for correspondence
%%\textsuperscript{number} is used for affiliation
%%\affilOne, \affilTwo etc., upto \affilTwentyfive is possible
%%Please note the first letter after \affil is capitalised in the command
%%

\author{J. Baghel\textsuperscript{1,*}, Silpa S.\textsuperscript{1}, P.  Kharb\textsuperscript{1}, B. Sebastian\textsuperscript{2} and P. Shastri\textsuperscript{3,**}}
\affilOne{\textsuperscript{1}National Centre for Radio Astrophysics - Tata Institute of Fundamental Research, S. P. Pune University Campus, Ganeshkhind, Pune 411007, India\\}
\affilTwo{\textsuperscript{2}Department of Physics \& Astronomy, Purdue University, 525 Northwestern Ave., West Lafayette, IN 47907, USA\\}
\affilThree{\textsuperscript{3}Indian Institute of Astrophysics, II Block Koramangala, Bengaluru 560034, India. \textsuperscript{**}(Retired) }

%%escape two column mode for title, affiliation and abstract
%%by giving \twocolumn command as shown
\onecolumn
\maketitle
%%include \corres to print the corresponding author Email id
\corres{jbaghel@ncra.tifr.res.in}

%%include \msinfo for
%%manuscript information such as
%%received, revised and accepted dates
%%
%\msinfo{1 January 2015}{1 January 2015}
%%abstract

\vspace{0.5cm}
\begin{abstract}
We present results from our radio polarimetric study with the upgraded Giant Metrewave Radio Telescope (uGMRT) at Band 4 ($550-850$~MHz) of 3 blazars: radio-loud quasars 3C390.3, 4C71.07 and BL Lac object 1ES~2344+514. The aim of this study was (i) to carry out a feasibility study for Band 4 polarization with the uGMRT, and (ii) to compare and contrast the kpc-scale polarization properties between the blazar sub-classes. We have detected linear polarization in all the three sources. 
{ The degree of linear polarization in the cores of the two quasars is higher than in the BL~Lac object, consistent with similar differences observed on parsec-scales in blazars.}
The highest fractional polarization of $\approx $15\% is observed in the hotspot region of 3C390.3, which also shows extended polarized lobe structures. 1ES~2344+514 shows a core-halo structure whereas 4C71.07 remains unresolved. A rotation of polarization electric vectors along the northern hotspot of 3C390.3, and the core of 1ES~2344+514, suggest jet bending. Greater depolarization in the southern lobe of 3C390.3 compared to the northern lobe indicates the presence of the `Laing-Garrington effect'. Multi-frequency uGMRT polarimetric data are underway to study the kpc-scale rotation measures across these sources in order to look for differences in the surrounding media.
\end{abstract}

%%insert keywords separated by 3 hyphens using \keywords{words}
%%close the twocolumn escape here
%%include \doinum{number}for the DOI number in the header
%%include \volnum{number} for the volume number in the header
%%include \year{yyyy} for  year of publication in the header
%%include \pgrange{num--num} page range of article in the header
%%include \artcitid{num} for the article citation id
%%include \lp to print last page of the article
%%include \setcounter{page}{pagenum} for the exact starting page of the article

\doinum{xyz/123}
\artcitid{\#\#\#\#}
\volnum{000}
\year{2021}
\pgrange{1--13}
\setcounter{page}{1}
\lp{13}

\section{Introduction}
Active galactic nuclei (AGN) are the highly luminous and energetic centres of a small fraction of galaxies. It is now widely believed that the extreme luminosities of AGN are a result of the release of gravitational energy as matter accretes onto supermassive black holes \citep[$\sim10^6-10^9$~M$_\odot$;][]{Rees1984}. \citet{Kellermann1989} have suggested an AGN division into the radio-loud (RL) and radio-quiet (RQ) categories based on the radio-loudness parameter, $R=\mathrm{S_{5~GHz}/S_{B-band}}$, where $\mathrm{S_{5~GHz}}$ is the radio flux density at 5~GHz and $\mathrm{S_{B-band}}$ is the (nuclear) optical B-band flux density. Radio-loud AGN have $R\gg10$ and are characterised by the presence of kpc-scale bipolar collimated relativistic jets launched perpendicular to their accretion disks. The RL-RQ divide is a longstanding open question in AGN physics. Differences in accretion rates, host galaxy types and environments have all been suggested in the literature to explain the different AGN types \citep[e.g.,][]{Prestage1988,Antonucci1993,Heckman2014,Padovani2016}.

Blazars are radio-loud AGN that are characterised by high luminosities, rapid variability, high and variable polarization, superluminal motion, and intense non-thermal emission across the electromagnetic spectrum. These extreme properties can be explained by Doppler beaming/boosting effects as blazars are suggested to have their jets oriented at small angles to our line of sight \citep{Urry1995}. Blazars have different optical emission line spectra: strong and broad (velocity widths of $\gtrsim 1000$~km~s$^{-1}$) emission lines are observed in flat spectrum radio quasars (FSRQs), while BL Lacertae objects (BL Lacs) are characterised by featureless spectra \citep[e.g.,][]{Stickel1991,Stocke1991}. This is the so called `blazar divide'. 
However, this dichotomy is still under dispute and multi-wavelength studies on these two classes reveal several other differences. On parsec scales, blazars have been extensively studied with very long baseline interferometry (VLBI) and it has been found that BL Lacs tend to have their parsec-scale polarization vectors parallel to the jet direction whereas FSRQs tend to show a perpendicular relative orientation \citep{Lister2013}. Also, BL Lacs have systematically higher parsec-scale rotation measures (RM) relative to the FSRQs \citep{Zavala2005}. 

We have carried out a pilot radio polarimetric study with the upgraded Giant Metrewave Radio Telescope (uGMRT) at Band 4 ($\sim600$~MHz) of 3 blazars: the radio-loud quasars 3C390.3, 4C71.07 and the BL Lac 1ES~2344+514. uGMRT polarimetric images can reveal the inferred kpc-scale magnetic (B) field structures in the jets and lobes, as they currently provide the best compromise between resolution and sensitivity at low radio frequencies. We have searched for differences in the polarization properties between the BL Lac and the quasars on kpc-scales by examining the polarization vector orientations and the degree of polarization, in their cores, jets and lobes. We present the results from our study here. Throughout this paper, we have adopted $\Lambda$CDM cosmology with $\mathrm{H_0 = 73~km~s^{-1} Mpc^{-1}}$, $\mathrm{\Omega_m = 0.27}$ and $\mathrm{\Omega_\nu = 0.73}$. The spectral index $\alpha$ is defined such that flux density at frequency $\nu$, $S_\nu \propto \nu^\alpha$.

\section{Sample Selection}
{ The selected quasars 3C390.3 ($z=0.0561$) and 4C71.07 ($z=2.172$) and the BL Lac 1ES~2344+514 ($z=0.044$), were chosen for multi-band quasi-simultaneous observations with the Whole Earth Blazar Telescope (WEBT) and ASTROSAT \citep[e.g.,][ASTROSAT PI: Prajval Shastri]{Villata00} as they represent the typical examples of their respective categories and have been extensively monitored at various frequencies in the literature. 3C390.3 is a bright, superluminal \citep{Porcas1987}, highly polarised \citep{Angel1980}, extended flat spectrum source \citep{Healey2007} identified as an optically violently variable (OVV) blazar \citep{Cannon1971} that has been extensively monitored over the years \citep{Ghosh2000}. Gamma-ray-loud blazars 1ES2344+514 and 4C71.07 are being monitored under the GLAST-AGILE Support Program (GASP) under the WEBT \citep{GASP2008}. } The three blazars span a wide range in redshift, total radio power, ratio of nucleus-to-host galaxy flux densities and the dominance of the inverse Compton (IC) peak in their spectral energy distributions (SED). The purpose of this study is to infer broad intrinsic trends between the blazar subclasses.

\section{Radio Data Analysis}
The uGMRT Band 4 ($550-850$ MHz)\footnote{The upper end of the band up to 900 MHz is affected by RFI from mobile phone band, URL:  {http://gmrt.ncra.tifr.res.in/doc/ugmrt.pdf}} full polarization mode observations of 3C390.3 and 1ES~2344+514 were carried out on 3 July 2017 and of 4C71.07 on 30 June 2017. Band 4 employs dipolar antenna feeds that receive dual linear polarization, which are then converted to circular polarization using a polarizer. The angular resolution at Band 4 is $\sim5''$ while the HPBW\footnote{Half Power Beam Width} of the primary beam is $43\pm3'$. For our experiment, 3C286 (polarized calibrator) was used as the primary flux calibrator as well as the polarization leakage and polarization angle calibrator. Several scans of 3C286 spread throughout the experiment, { ensured} a good parallactic angle coverage for a good leakage calibration. 2355+498, 2005+778 and 0921+622 were used as the phase calibrators and 3C48 as the secondary flux calibrator. A single scan of OQ208 (unpolarized calibrator) was also included; it was however finally not used to calculate the instrumental leakage terms due to its faintness in Band 4 ($\sim$0.5~Jy), which could not provide sufficient signal-to-noise ratio (SNR) to accurately determine the instrumental polarization for the uGMRT. 3C84 (unpolarized calibrator) was used for leakage calibration for observations carried out on 3 July 2017. 

\subsection{Data Reduction and Polarization Pipeline}
A CASA\footnote{Common Astronomy Software Applications; \citet{mcmullin2007}} based polarization pipeline developed by \citet{Silpa2020}\footnote{A Python script for the polarization pipeline of uGMRT data is available at https://sites.google.com/view/silpasasikumar/} was used for the data analysis \citep[see also][]{Silpa2021}. We briefly discuss the data reduction steps below. We first converted the LTA data files produced by the uGMRT to FITS format using LISTSCAN and GVFITS utilities. We restricted the frequency range of our data to $\sim 580-660$~MHz by editing the log file produced from LISTSCAN. We also omitted the non-working antennas in the log file. The frequency range was decided upon two factors: (i) the sensitivity for uGMRT Band 4 data drops significantly beyond $560-810$~MHz range; (ii) the uGMRT was undergoing an upgradation during the period of these observations and not all antenna feeds were upgraded at that time. Therefore, we carried out a test run on the full bandwidth data and identified the maximum possible frequency range that yielded a flat bandpass gain for most antennas. The FITS file was then converted to measurement set (MS) using the task IMPORTUVFITS in CASA. Flagging of RFI-affected data was carried out at various stages using TFCROP and RFLAG modes of CASA task FLAGDATA. 

{ The task SETJY in CASA was used to set the flux density values for the amplitude calibrators. The output of SETJY was 20~Jy for 3C286 (using the Perley-Butler 2017 scale) and 26~Jy for 3C48.} This was followed by basic calibration steps involving initial phase calibration, bandpass and delay calibration, and phase and amplitude calibration. While bandpass and delay calibration were carried out using the full frequency range, the gain calibrations were carried out using the central 75\% good channel range. The task SETJY in CASA was used to set the model of a polarized calibrator by providing parameters like reference frequency, Stokes I flux density value at the reference frequency, spectral index value, and Taylor expansion coefficients for fractional polarization and polarization angle by fitting a first-order polynomial to their values obtained from the NRAO VLA observing guide\footnote{https://science.nrao.edu/facilities/vla/docs/manuals/obsguide/ modes/pol} as a function of frequency about the reference frequency. This was followed by polarization calibration steps which involved cross-hand delay calibration using 3C286, leakage calibration (`D-terms') using 3C84 for the 3 July 2017 observations and 3C286 for the 30 June 2017 observations, and polarization angle calibration using 3C286. The D-term amplitudes turned out to be typically $\lesssim20-25$\% for the uGMRT antennas.

{We note that the inferred B-field vectors have not been corrected for Faraday rotation. There are currently no widely available models to correct for differences in the observing geometry or the varying magnetic field strength in the ionosphere for the GMRT antennas.} Using the total electron content (TEC) maps alone, \citet{Farnes2014} noted that the maximum Faraday rotation due to the ionosphere was $\leq2$~rad~m$^{-2}$ at 610~MHz for the legacy GMRT for the duration of their observations (3-4 hours on source). We plan to obtain more refined estimates for the upgraded GMRT in the near future. Ionospheric corrections of the order of $0.1-0.3$~rad~m$^{-2}$ have been noted for LOFAR observations at 150~MHz by \citet{Mahatma2021}. Most importantly, the results of our uGMRT observations match those from the WSRT where ionspheric corrections were included \citep{Jaegers1987}, to within 5-10 degrees, giving us confidence about the correctness of our findings. We also stress that our aim here is to study the B-field structures in the kpc-scale jets/lobes of blazars, rather than their cores, where the RM is likely to be higher \citep[e.g.,][]{Saikia1987}. The average kpc-scale RM values, as would be relevant for the large jets/lobes of these sources, have been estimated for 3C390.3 to be $-5\pm1$~rad~m$^{-2}$ by \citet{SimardNormandin81}, and $-11\pm7$~rad~m$^{-2}$ for 4C71.07 by \citet{Wrobel93}.
{However, we note that RM values can be higher than the integrated RM values at specific locations along the source length, which can influence the local inferred B-field directions
\citep[e.g.,][]{Dreher1987,McKean2016,Silpa2021B}.}

The flux density values for phase calibrators were determined using the task FLUXSCALE in CASA. The calibration solutions were applied to the multi-source datasets and visibility data for individual targets were extracted using the CASA task SPLIT while averaging the spectral channels to account for bandwidth smearing effects. The Stokes I images of our sources were created using the multiterm-multifrequency synthesis \citep[MTMFS;][]{Rau2011} algorithm in CASA task TCLEAN. This was followed by four rounds of phase-only self-calibration and five rounds of amplitude and phase self-calibration. The Stokes Q and U images were created using the final self-calibrated visibility data. We used same imaging parameters as for the Stokes I image except for lesser number of iterations. 

The AIPS\footnote{Astronomical Imaging Processing System; \citet{Wells1985}} task COMB was used to create the polarized intensity ($P = \sqrt{Q^2 + U^2}$; PPOL) and electric vector polarization angle (EVPA or $\chi = 0.5~tan^{-1} (U/Q)$; PANG) images from Stokes Q and U images by setting the parameter opcode = POLC (which corrects for Ricean bias) and POLA respectively. Regions with intensity values below 3 times the rms noise and angle errors $>$ $10^\circ$ were blanked while making the PPOL and PANG images respectively. The task COMB with opcode = DIV was used to create fractional polarization ($FP= P/I; FPOL$) images while blanking regions with fractional polarization errors $>$10\%. The Gaussian-fitting AIPS task JMFIT was used to estimate flux densities for compact components whereas the AIPS task TVSTAT was used for extended emission. The distances reported in the paper were measured using the AIPS task TVDIST.

Certain modifications were made to the original pipeline. The phase referencing antenna mode (refantmode) `flex' was changed to `strict'. The `flex' refantmode switches to an alternate reference antenna if the preferred or current reference antenna drops out. If and when the preferred reference antenna returns, the referencing switches back to it. The original, preferred reference antenna may now be at a non-zero phase difference with respect to the alternate reference antenna. It will return a new phase value that will be kept constant from that point on. Under low SNR conditions, the effective cross-hand phase is unstable at reference antenna changes, since each hand of polarization is referenced independently, and the non-zero noise on the cross-hand phase difference will change. When carrying out polarimetric data reduction (which depend on stable cross-hand phases), it is preferable to use the refantmode `strict' which keeps only one reference antenna and if it is absent for a solution, flags all antennas at that solution. Other modifications made included removing syntax discrepancies between different CASA versions (5.5.0 vs 5.6.1) used in developing the original pipeline and the one used in this analysis.

\begin{figure}[t]
\centering
  \includegraphics[width=6.9cm,trim=42 0 0 20]{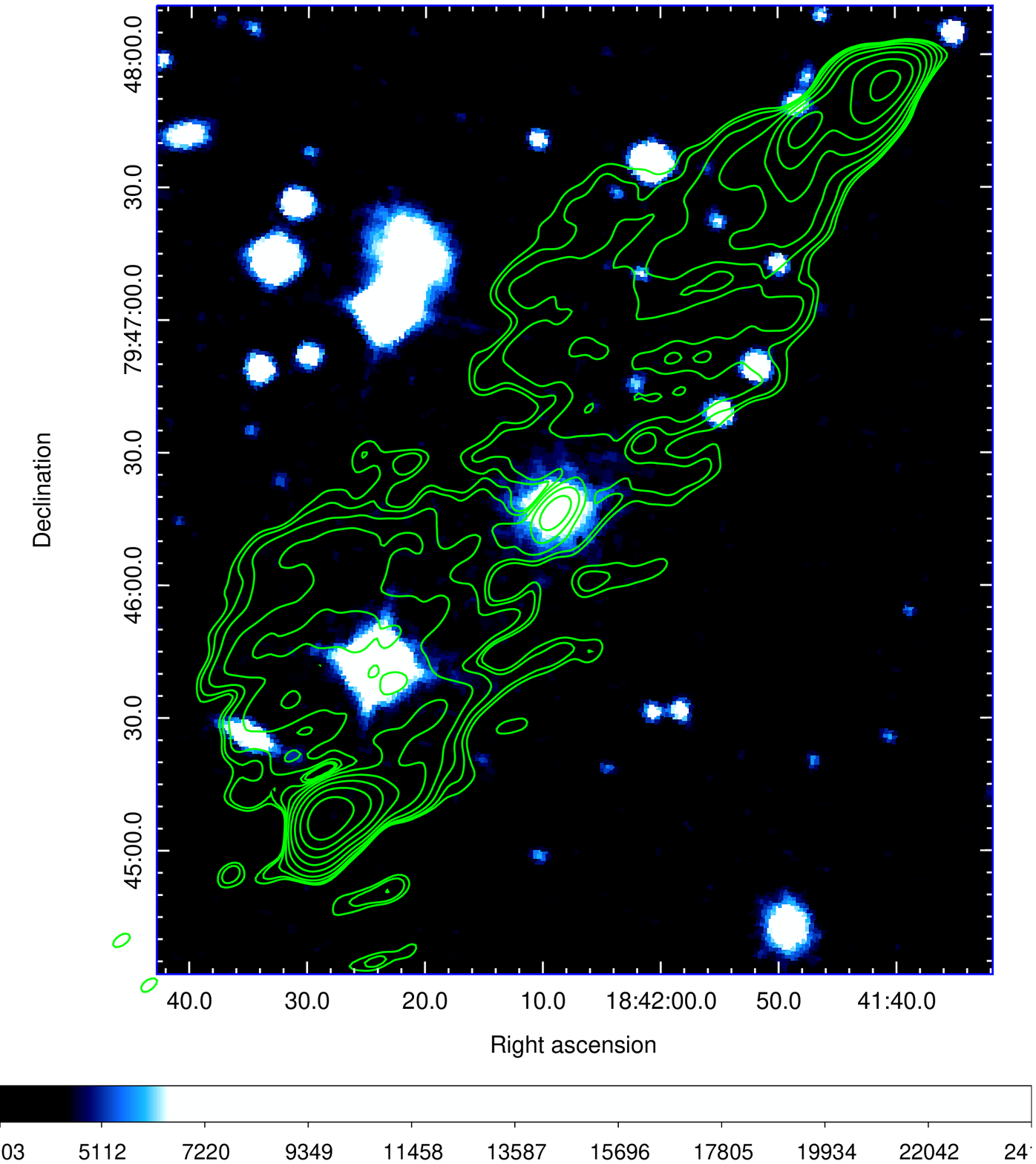}
  \includegraphics[width=8.7cm,trim=0 0 0 30]{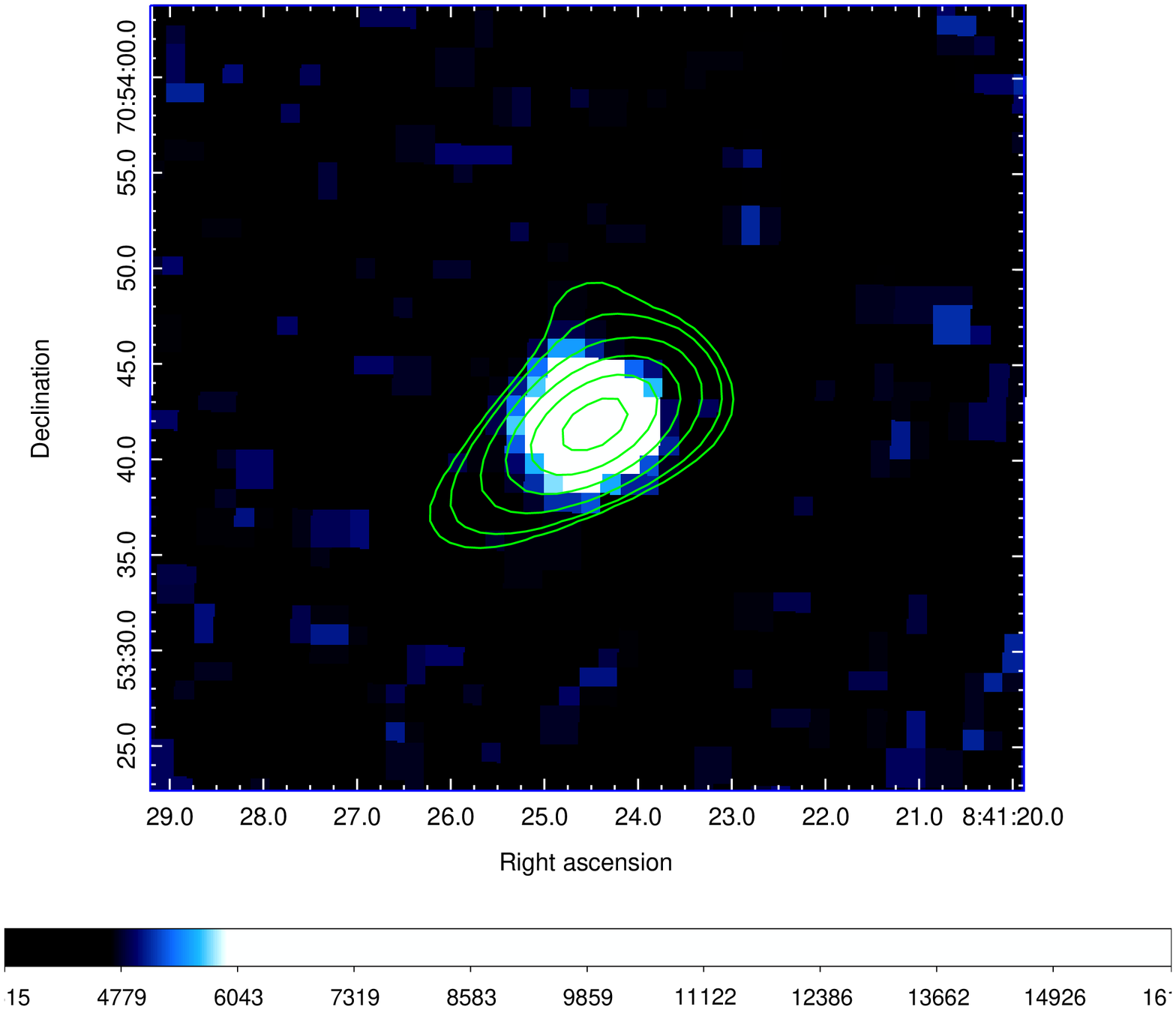}
   \includegraphics[width=11cm]{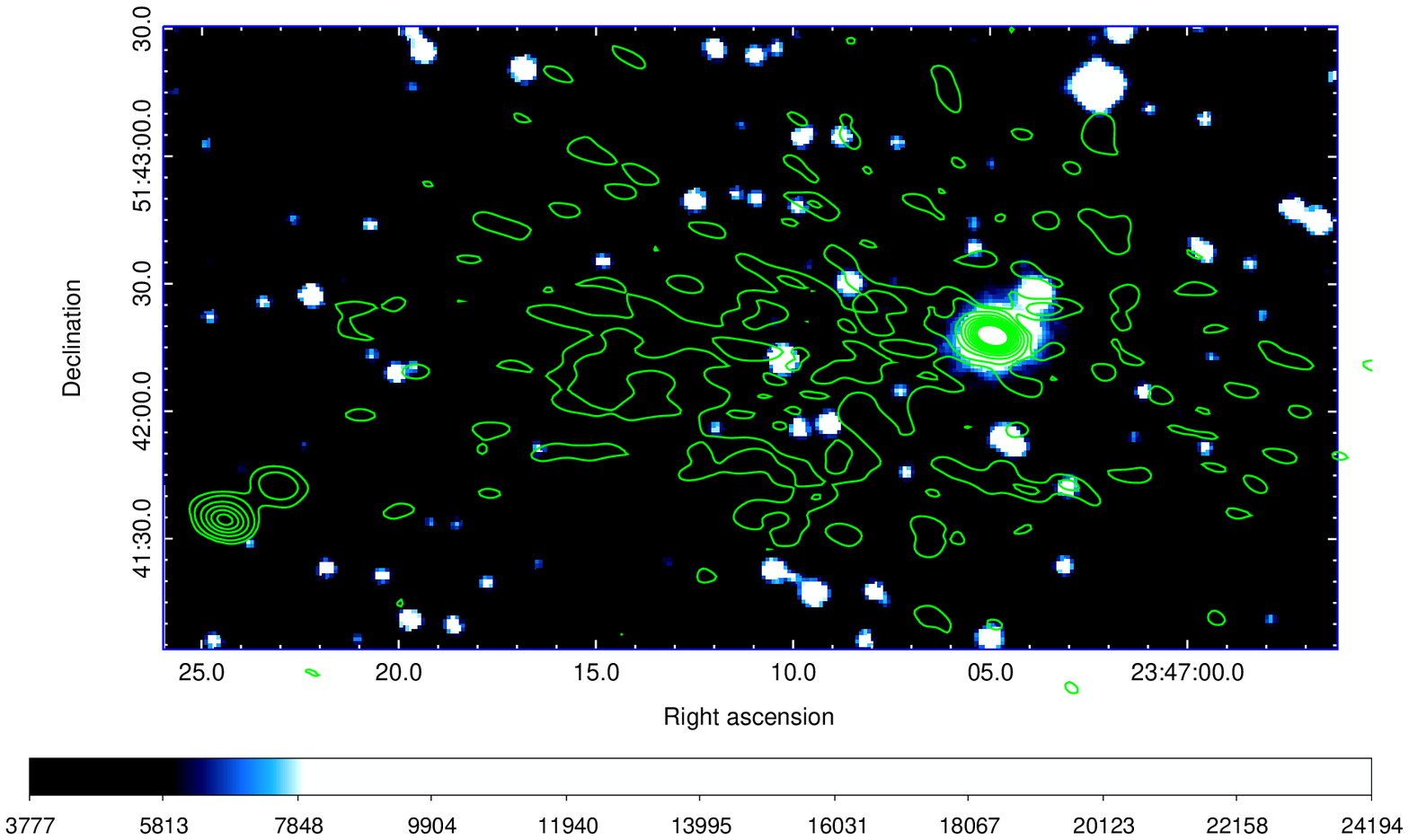}
  \caption{The radio-optical overlays of (top left) quasar 3C390.3, (top right) quasar 4C71.07, and (bottom) 1ES~2344+514, using DSS optical images and uGMRT Band 4 radio contours. The elliptical host galaxies of the 3 blazars are clearly visible. }
  \label{fig:radio_optical}
\end{figure}

\section{Results and Discussion}
\begin{figure}
\centering
\includegraphics[width=14cm, trim = 0 70 0 30]{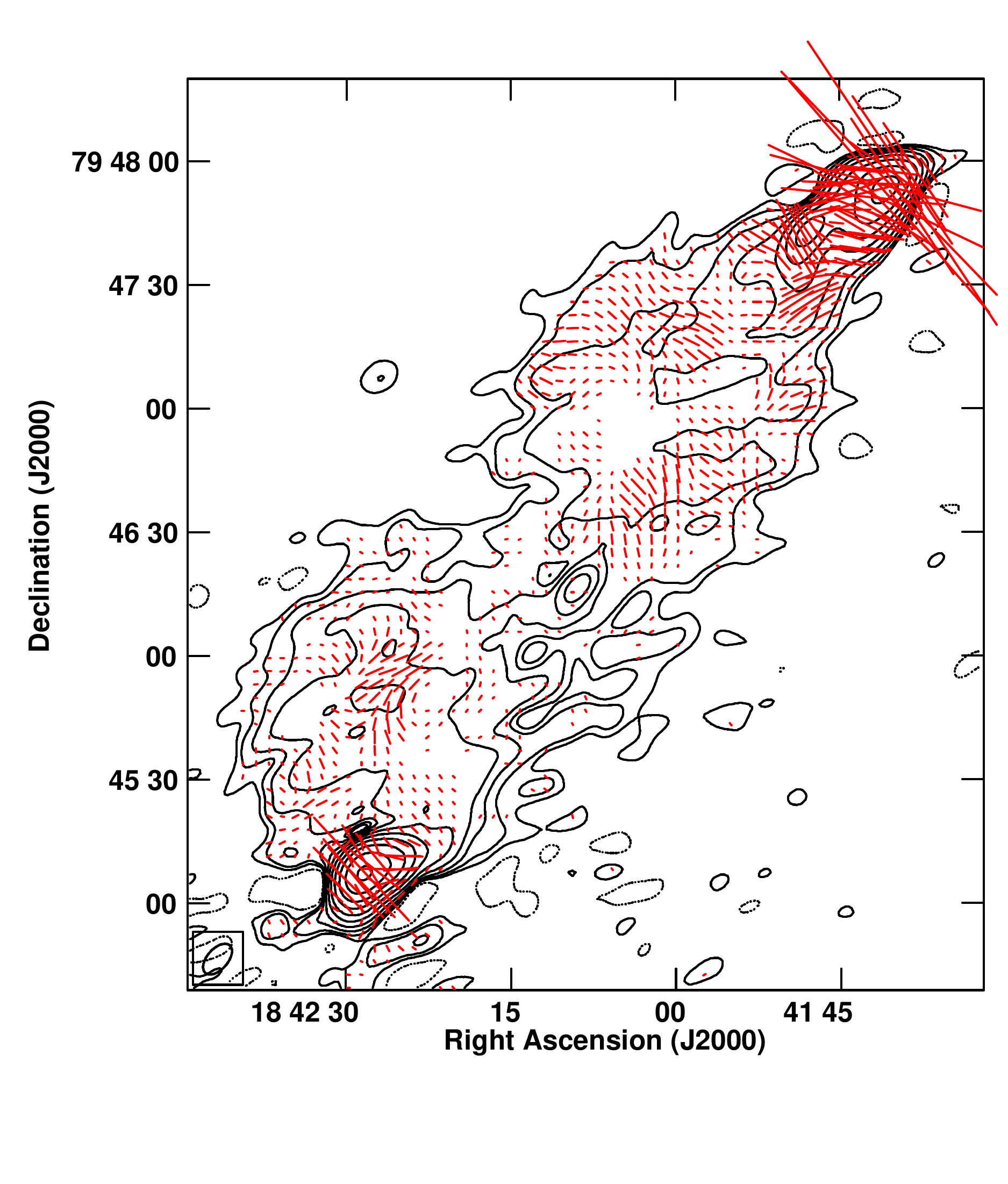}
%\caption{{The 620 MHz contour image of the quasar 3C390.3 with EVPA vectors superimposed as red ticks. The beam is $8.67'' \times 4.99''$ with a PA of $-43.17^\circ$. The peak surface brightness, $I_P$ is 4.16~Jy~beam$^{-1}$. The contour levels in percentage of the peak surface brightness $I_P$ are $(-0.18,~0.18,~0.35,~0.7,~1.4,~2.8,~5.6,~11.25,~22.5,~45,~90)$~Jy~beam$^{-1}$. The length of the EVPA vectors is proportional to polarized intensity with $20''$ corresponding to 62.5~mJy~beam$^{-1}$.}}
\caption{{The 620 MHz contour image of the quasar 3C390.3 with EVPA vectors superimposed as red ticks. The beam is $8.93'' \times 5.00''$ with a PA of $-41.45^\circ$. The peak surface brightness, $I_P$ is 3.92~Jy~beam$^{-1}$. The contour levels in percentage of the peak surface brightness $I_P$ are $(\pm 0.18,~0.35,~0.7,~1.4,~2.8,~5.6,~11.25,~22.5,~45,~90)$~Jy~beam$^{-1}$. The length of the EVPA vectors is proportional to polarized intensity with $20''$ corresponding to 41.6~mJy~beam$^{-1}$.}}
\label{fig:3C390}
\end{figure}

We have detected linear polarization in the cores/lobes of all the 3 blazars at $\sim$600~MHz. 3C390.3 exhibits the most extensive polarization structure in its core, lobes and hotspots (see Figure \ref{fig:3C390}), while polarization is primarily detected in the cores of 1ES~2344+514 and 4C71.07 (see Figures \ref{fig:1ES2344z} and \ref{fig:4C71.07}). We discuss the individual sources ahead. Table \ref{table1} lists the values of polarized intensity, total intensity, and fractional polarization for the different regions of the blazars. 

\begin{figure}
\centering
\includegraphics[width=14cm, trim = 0 70 0 30]{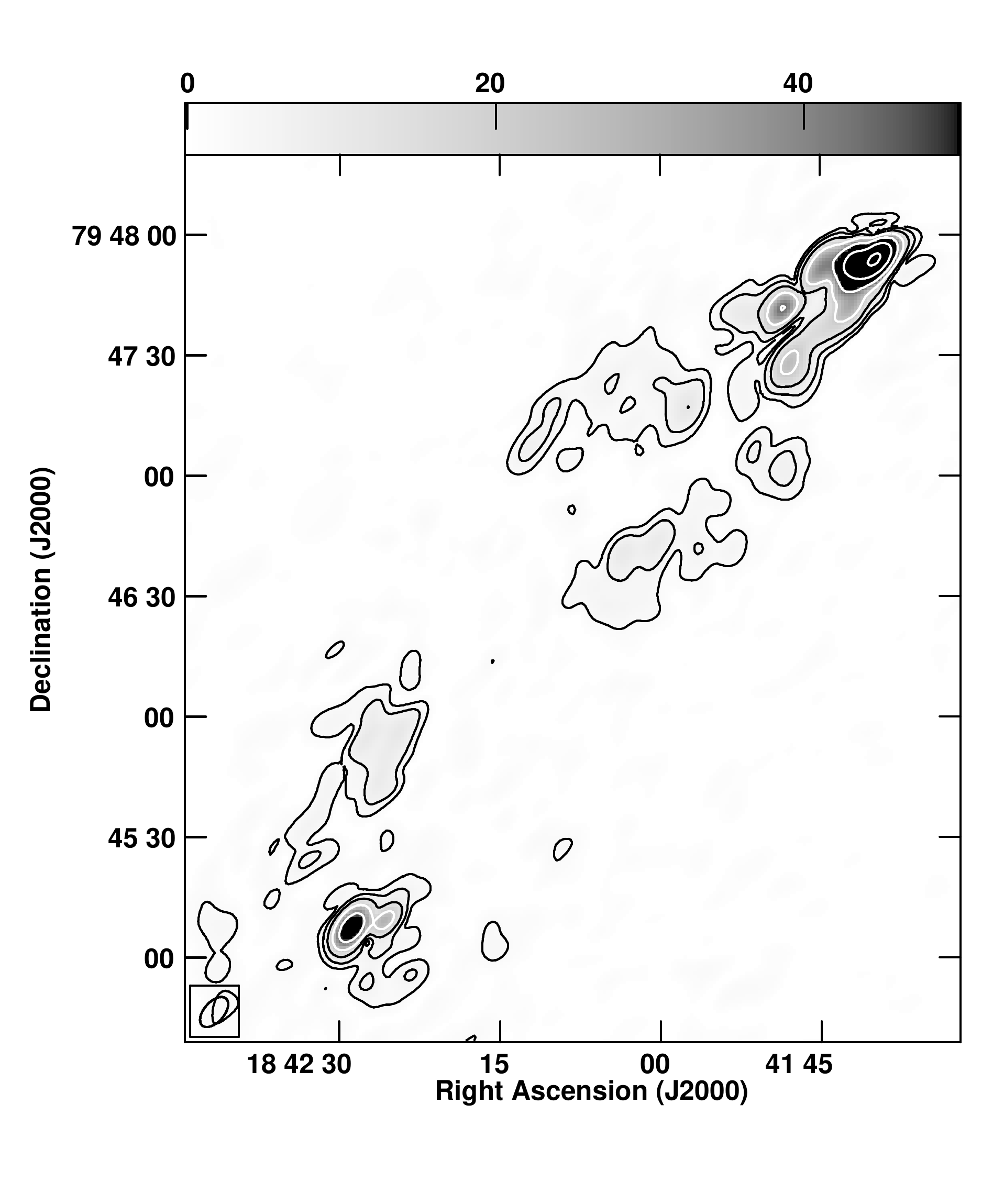}
%\caption{620 MHz polarized intensity in contours and greyscale with units in mJy~beam$^{-1}$. The peak polarised surface brightness ($P_P$) is 0.20~Jy~beam$^{-1}$. The contour levels in percentage of the peak polarised surface brightness $P_P$ are $(1.4,~2.8,~5.6,~11.25,~22.5,~45,~90)$~Jy~beam$^{-1}$.}
\caption{620 MHz polarized intensity in contours and greyscale with units in mJy~beam$^{-1}$. The peak polarised surface brightness ($P_P$) is 0.20~Jy~beam$^{-1}$. The contour levels in percentage of the peak polarised surface brightness $P_P$ are $(1.4,~2.8,~5.6,~11.25,~22.5,~45,~90)$~Jy~beam$^{-1}$.}
\label{fig:3C390_POL}
\end{figure}

{\begin{table}

\begin{center}

\tabularfont

\caption{Total and polarized intensity estimates for the 3 blazars}\label{table1} %%10/12

\begin{tabular}{lcllll}

\hline

Source & Region & & P~(mJy)& I~(mJy) & FP~(\%) \\\hline

3C390.3 & Core : & & {1.4 $\pm$ 0.3} & {231} & { 1.2 $\pm$ 0.3 } \\

& Hotspot region : & North-west & { 483 $\pm$ 5} & { 4.0E+03} & { 12 $\pm$ 0.4} \\

& & South-east & { 131 $\pm$ 4}& { 8.0E+03} &{ 3.6$\pm$ 0.4 }\\
	
& Lobes : & North-west & { 342 $\pm$ 32}& { 3.4E+03} & { 14 $\pm$ 2} \\

&& South-east & { 196 $\pm$ 25}& { 4.7E+03} & { 7.2 $\pm$ 2} \\\hline

1ES~2344+514 & Core : & & { 0.99 $\pm$ 0.08} & { 278} & { 0.63 $\pm$ 0.08 } \\

 & Eastern source :& & { 0.4  $\pm$ 0.1} & { 59.8 }& { 9 $\pm$ 3} \\\hline

4C71.07 & Core : & &{ 129.8 $\pm$ 0.9} & { 4.6E+03 } & { 5.56 $\pm$ 0.06} \\ \hline

\end{tabular}

\tablenotes{P = Polarized Intensity, I = Total Intensity, FP = Fractional Polarization }

\end{center}

\end{table}

}

\subsection{Quasar 3C390.3}
\begin{figure}
\centering
\includegraphics[width=8.9cm, trim = 0 35 0 200]{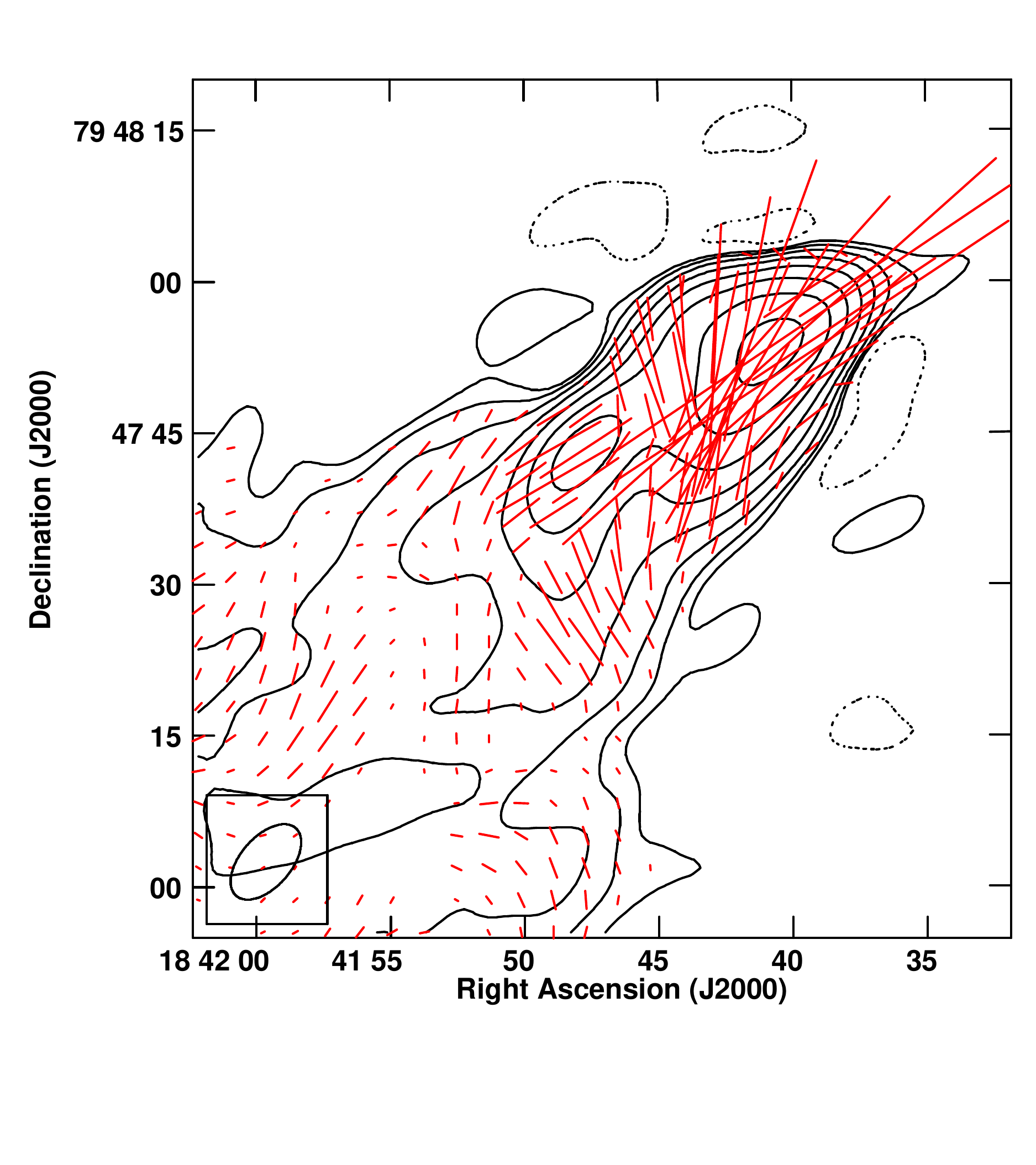}
\includegraphics[width=8.3cm]{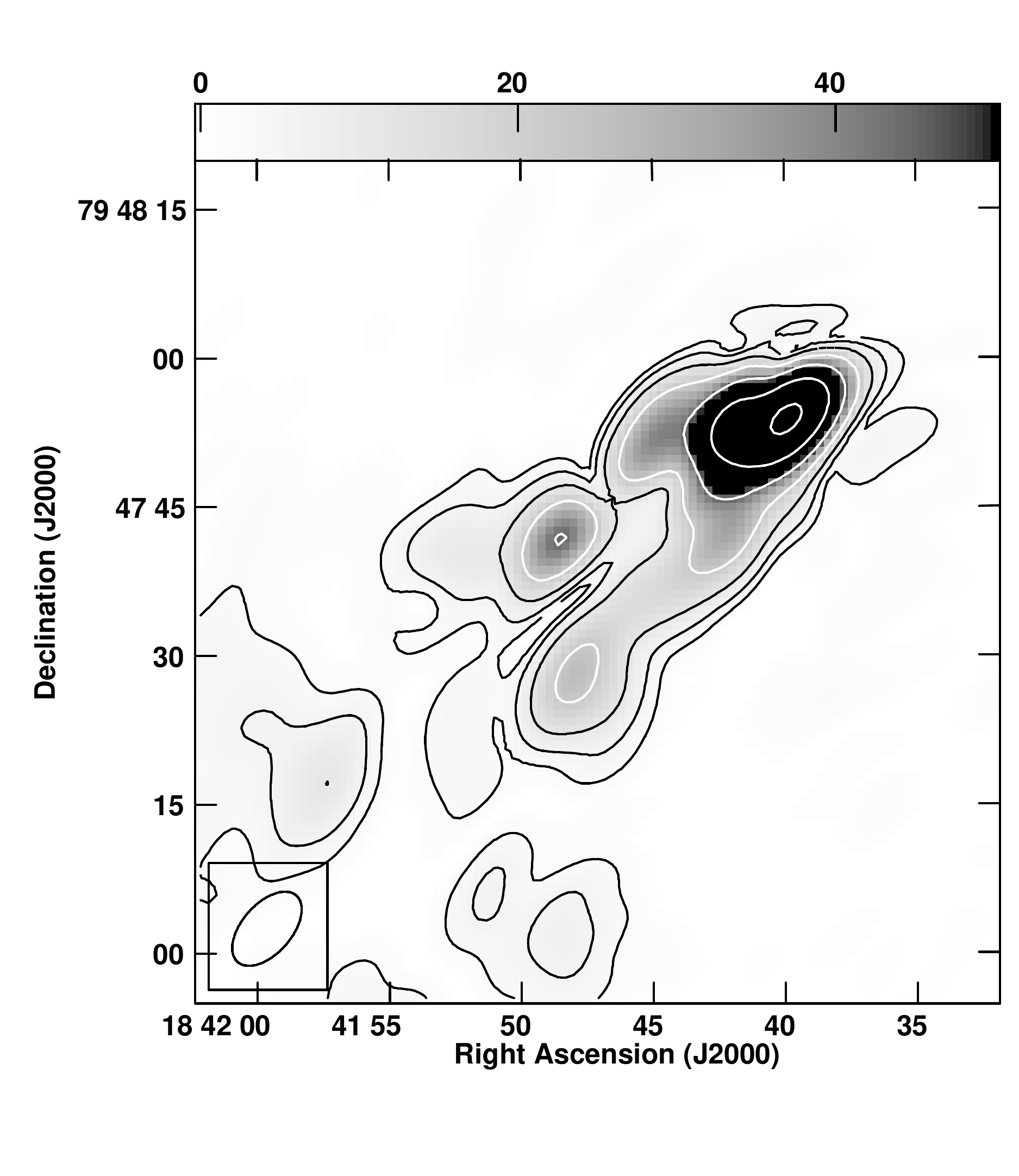}
%\caption{{ (Left) 620~MHz radio contours with B-field vectors (EVPAs rotated by $90^\circ$ assuming optically thin emission) zoomed-in on the northern hotspot of the quasar 3C390.3. The beam is $8.67'' \times 4.99''$ with a PA of $-43.17^\circ$. The peak surface brightness ($I_P$) is 4.16~Jy~beam$^{-1}$. The contour levels in percentage of the peak surface brightness $I_P$ are $(\pm0.35,~0.7,~1.4,~2.8,~5.6,~11.25,~22.5,~45,~90)$~Jy~beam$^{-1}$. The length of the EVPA vectors is proportional to polarized intensity with $5''$ corresponding to 20.8~mJy~beam$^{-1}$.(Right) 620 MHz polarized intensity in contours and greyscale zoomed-in on the northern hotspot of the quasar 3C390.3 with units in mJy~beam$^{-1}$. The peak polarised surface brightness ($P_P$) is 0.20~Jy~beam$^{-1}$. The contour levels in percentage of the peak polarised surface brightness $P_P$ are $(1.4,~2.8,~5.6,~11.25,~22.5,~45,~90)$~Jy~beam$^{-1}$. }}
\caption{{ (Left) 620~MHz radio contours with B-field vectors (EVPAs rotated by $90^\circ$ assuming optically thin emission) zoomed-in on the northern hotspot of the quasar 3C390.3. The beam is $8.93'' \times 5.00''$ with a PA of $-41.45^\circ$. The peak surface brightness, $I_P$ is 3.92~Jy~beam$^{-1}$. The contour levels in percentage of the peak surface brightness $I_P$ are $(\pm0.18,~0.35,~0.7,~1.4,~2.8,~5.6,~11.25,~22.5,~45,~90)$~Jy~beam$^{-1}$. The length of the EVPA vectors is proportional to polarized intensity with $5''$ corresponding to 15.6~mJy~beam$^{-1}$.(Right) 620 MHz polarized intensity in contours and greyscale zoomed-in on the northern hotspot of the quasar 3C390.3 with units in mJy~beam$^{-1}$. The peak polarised surface brightness ($P_P$) is 0.20~Jy~beam$^{-1}$. The contour levels in percentage of the peak polarised surface brightness $P_P$ are $(1.4,~2.8,~5.6,~11.25,~22.5,~45,~90)$~Jy~beam$^{-1}$.}}
\label{fig:3C390_B}
\end{figure}

Figure~\ref{fig:radio_optical} shows the radio-optical overlay of 3C390.3 using the DSS optical image and the uGMRT 620 MHz contours. The elliptical host galaxy is clearly seen in this image. The 620 MHz image of 3C390.3 is presented in Figure~\ref{fig:3C390}. Here the total intensity radio contours are superimposed on by EVPA vectors in red. The synthesized beam is $8.93'' \times 5.00''$ with a position angle (PA) of  $-41.45^\circ$  %The rms noise in the total intensity image is { 3100}~$\mu$Jy~beam$^{-1}$ and polarized intensity image is { 365}~ $\mu$Jy~beam$^{-1}$. 
The rms noise in the total intensity image is { 2080}~$\mu$Jy~beam$^{-1}$ and polarized intensity image is { 220}~ $\mu$Jy~beam$^{-1}$. 
We observe a typical double radio source with lobes and terminal hotspots in 3C390.3. Overall, the fractional polarization in the north-western lobe and hotspot is higher than that in its south-western lobe and hotspot. Transverse EVPA vectors were seen in a jet segment entering the northern hotspot implying parallel B-fields, assuming optically thin emission. These could represent the poloidal B-field component associated with the radio jet. The EVPA vectors then appear to rotate in the hotspot region. This could suggest that the B-fields are compressed due to shocks in the hotspot region \citep{Gabuzda1992,Laing1980} (see Figure \ref{fig:3C390_B}). The EVPA rotation further implies a rotation of the jet head itself. 
\citet{Krause2018} have suggested that if a supermassive binary black hole system possesses a radio jet, it will show a long-term precession, which can potentially be observed as morphological features in radio images. Along with the wide terminal hotspot and jet-lobe misalignment (hotspots are seen at one edge of the lobe in 3C390.3 rather than in the centre), the rotation of B-field vectors along the northern hotspot in our image is consistent with jet precession. This would be consistent with the results found by \citet{Krause2018} in their 5 GHz image for 3C390.3 and may support the binary black hole hypothesis for it. Jet precession could however also occur due to accretion disk warping due to radiation instabilities or other effects \citep[see][]{Kharb06}. Alternatively, our image could be consistent with jet bending, possibly due to strong jet-medium interaction.

The southern lobe and hotspot show a polarization structure typical of double-lobed sources \citep[e.g.,][]{Kharb08} with the B-fields being perpendicular to the jet in the hotspot. The southern lobe and hotspot also show greater depolarization compared to the northern one. This is consistent with the northern jet/lobe being the approaching side and the southern lobe emission passing through a greater depolarizing medium, consistent with the `Laing-Garrington effect' \citep{Laing1988,Garrington1988}. Indeed, the presence of copious amounts of hot X-ray emitting gas has been observed surrounding the lobes of 3C390.3 in Chandra observations \citep{Kadler2004}. This is also consistent with the presence of the one-sided parsec-scale jet pointing towards the north-western lobe in the 15~GHz VLBI image of \citet{Kellermann2004}.

We note that we also created a polarization image at 610 MHz with a circular synthesised beam of $29''$ to match the published 610 MHz WSRT polarization image of 3C390.3 in \citet{Jaegers1987}. We found similar values of total fractional polarization ($\sim$ 2\%) and matching EVPA trends for the various components of the source like the hotspots and lobes; in fact, five distinct regions of polarization structures are similarly seen in both the WSRT and uGMRT images.

\subsection{BL Lac 1ES~2344+514}
A radio-optical overlay of the BL Lac object 1ES~2344+514 is presented in Figure \ref{fig:radio_optical} and clearly shows the elliptical host galaxy. The DSS optical image in colour is superimposed by uGMRT 635 MHz contours. The 635 MHz image of 1ES~2344+514 is presented in Figure~4. The total intensity radio contours are superimposed by EVPA vectors in red. The beam is $6.40''\times 3.99''$ at a PA of $75^\circ$. The rms noise in the total intensity image is { 227} $\mu$Jy~beam$^{-1}$ and polarized intensity image is { 21} $\mu$Jy~beam$^{-1}$. A core-halo structure is observed with a diffuse radio halo extending towards east from the source. The polarization structure in the core is complex but could be consistent with a jet bend smaller than the uGMRT beam. 

\begin{figure} 
  \centering
  \includegraphics[width=9.4cm,trim=0 0 0 100]{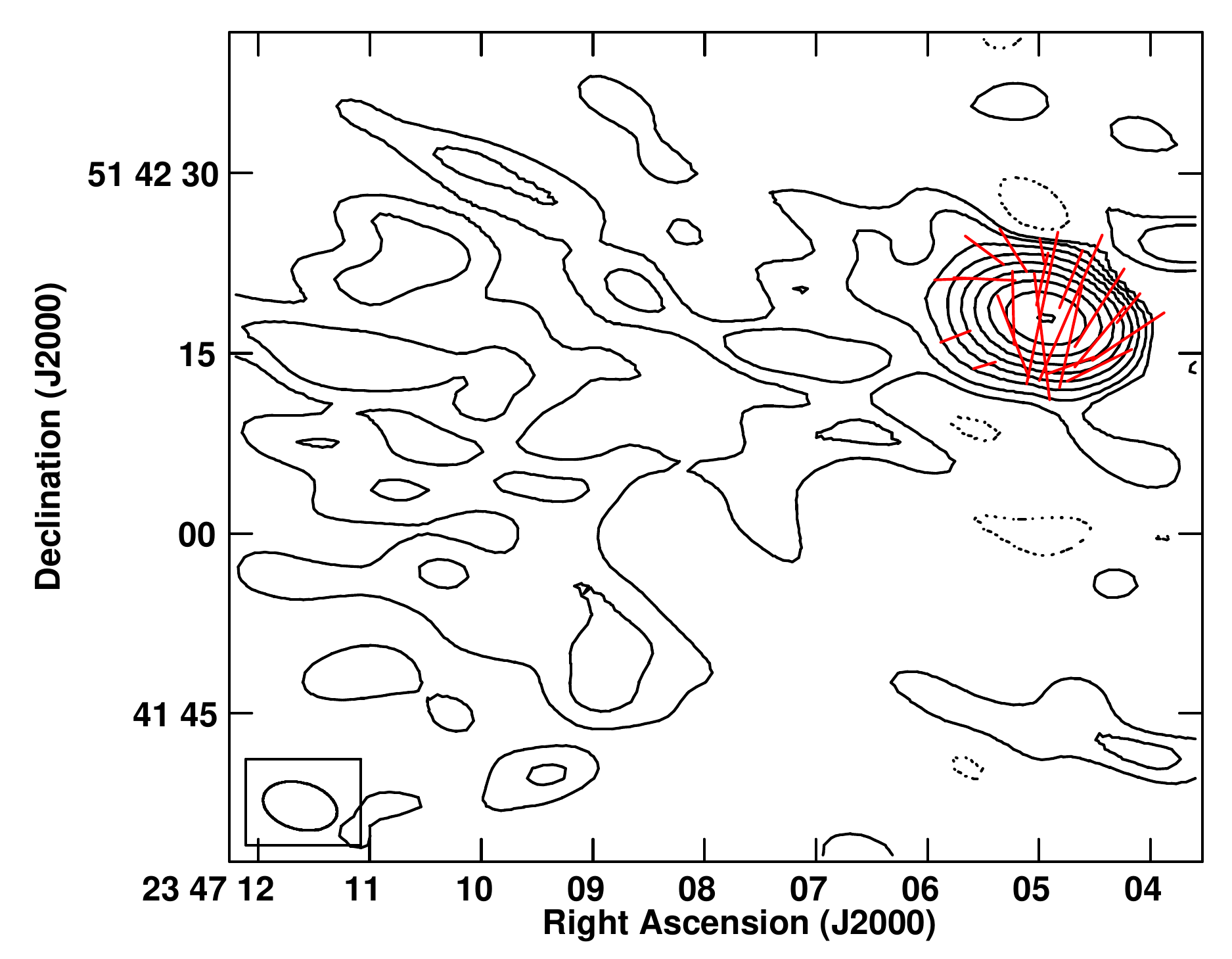}
  \includegraphics[width=7.5cm]{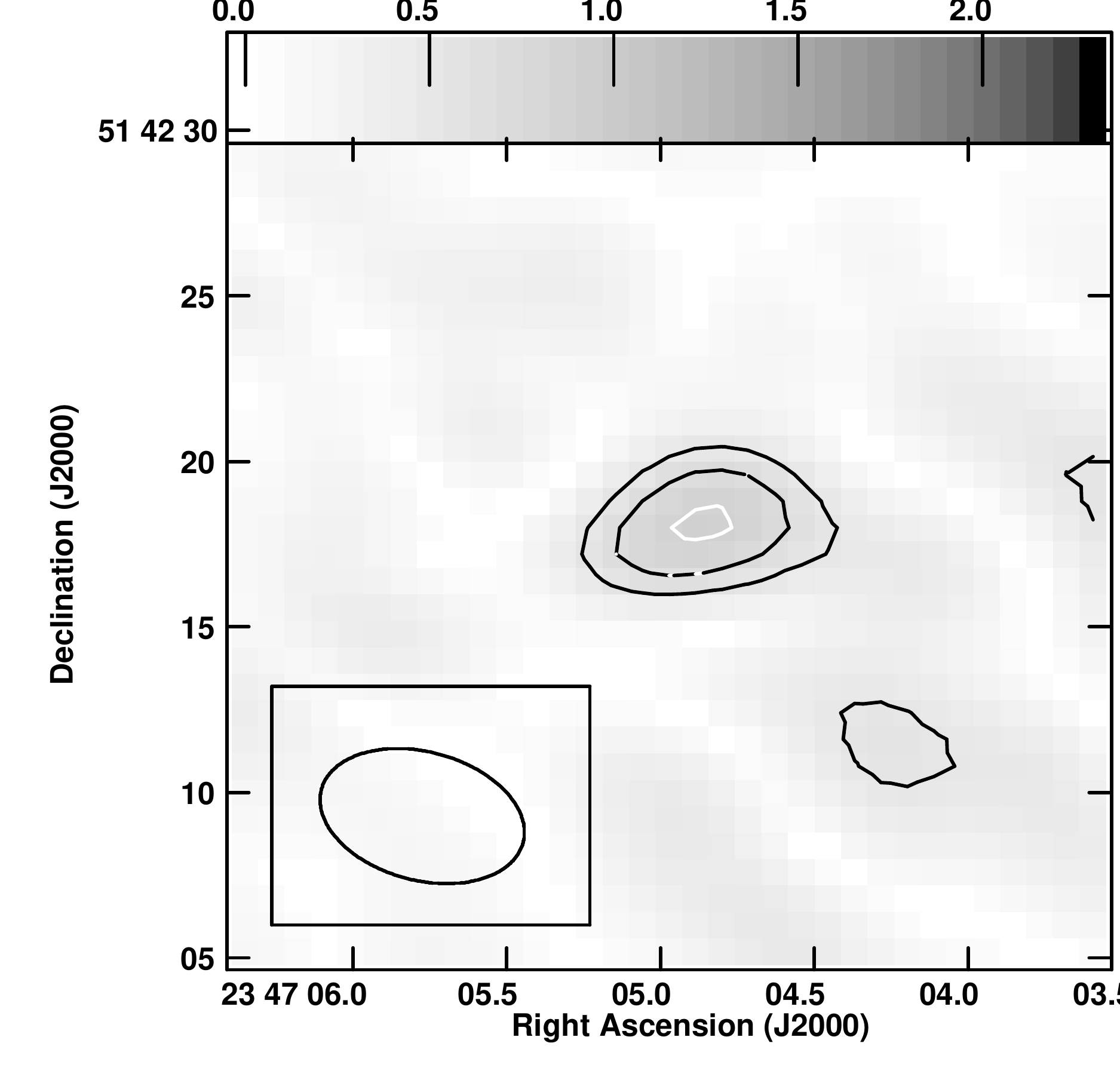}
  \includegraphics[width=8.7cm]{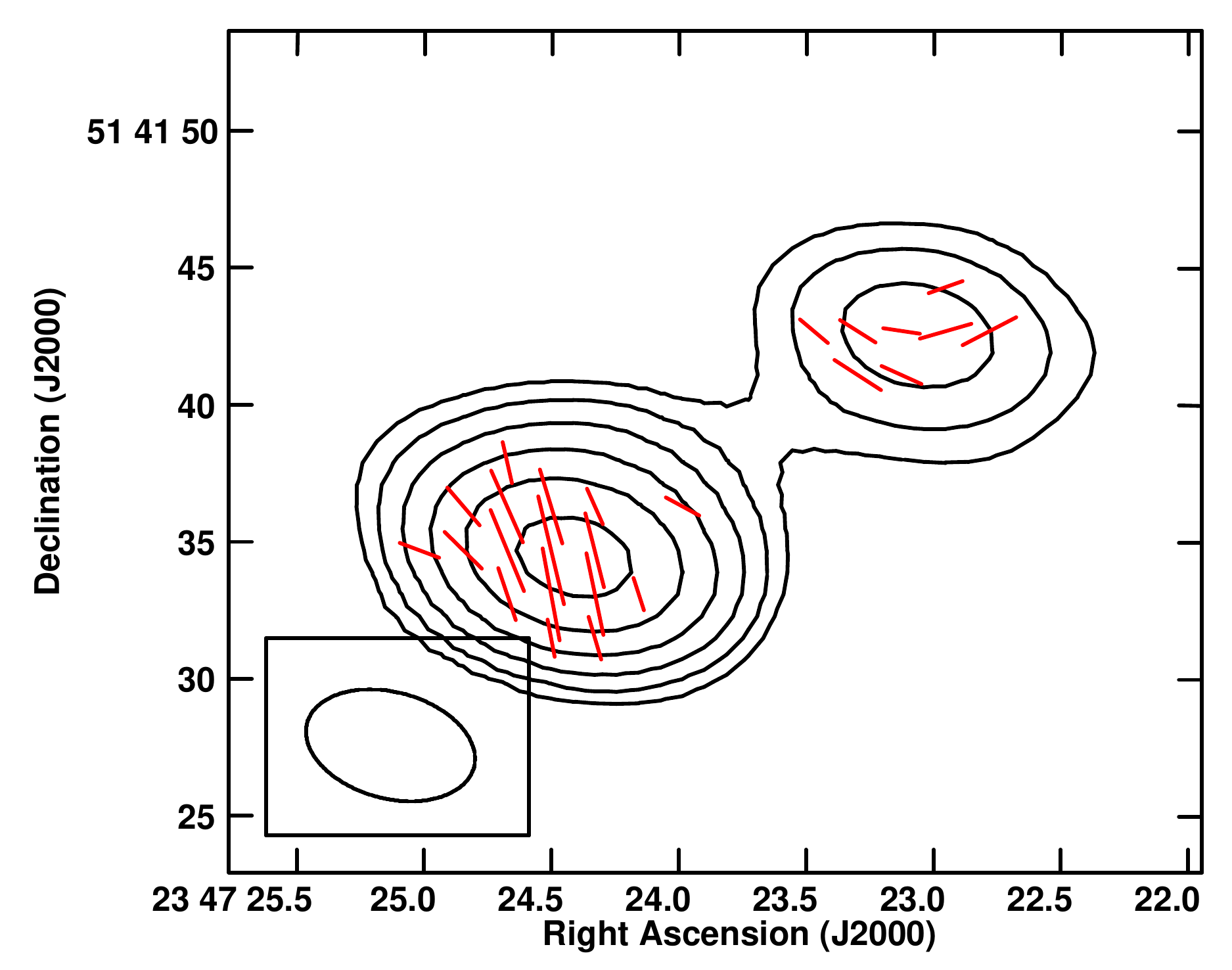}
  \caption{{ (Top Left) 635 MHz contour images of the BL Lac object 1ES~2344+514 with B-field vectors (EVPA vectors rotated by $90^\circ$) zoomed in on the core-halo and (Bottom) the Eastern source near it. The beam is $6.40''\times3.99''$ at a PA of $75^\circ$. The peak surface brightness ($I_P$) is 0.52~Jy~beam$^{-1}$. The contour levels in percentage of the peak surface brightness $I_P$ are $(\pm0.18,~0.35,~0.7,~1.4,~2.8,~5.6,~11.25,~22.5,~45,~90)$~Jy~beam$^{-1}$ for the top left panel and $(0.18,~0.35,~0.7,~1.4,~2.8,~5.6)$~Jy~beam$^{-1}$ for the bottom panel. The length of the EVPA vectors is proportional to polarized intensity with $5''$ corresponding to 0.31~mJy~beam$^{-1}$ for the top left panel and $2''$ corresponding to 0.13~mJy~beam$^{-1}$ for the bottom panel. (Top right) The polarized intensity in contours and greyscale with units in $\mu$Jy~beam$^{-1}$ of the core of 1ES~2344+514. The peak polarised surface brightness ($P_P$) is 9.29~mJy~beam$^{-1}$. The contour levels in percentage of the peak polarised surface brightness $P_P$ are $(5.5,~7.5,~9.5)$~Jy~beam$^{-1}$. }}
  \label{fig:1ES2344z}
\end{figure}

The 5~GHz parsec-scale Very Long Baseline Array (VLBA) image of 1ES~2344+514 by \citet{Bondi2004} shows a narrow one-sided jet which extends to about 20 parsec from the compact core with a position angle of $145^\circ$, and then flares and bends towards the south. Our uGMRT image shows that the kpc-scale jet is eventually bent towards the east. In order to check if the jet flaring in 1ES~2344+514 is associated with some intrinsic property of the central black hole, we estimated the `Bondi radius' for this source. Bondi accretion is spherical accretion onto a compact object like a black hole, with the Bondi radius ($r_B$) being the radius where the escape velocity equals the sound speed \citep{Bondi1952}. The gravitational potential changes abruptly and gas accretion changes from subsonic to supersonic infall. The jet collimates (and accelerates) in a semi-parabolic fashion steadily from the black hole to the Bondi radius where there is a change in the slope or the jet collimation break. Beyond that, the jet flares in a linear fashion and decelerates to sub-relativistic speeds \citep{Asada2012,Blandford2019}. $r_B$ was estimated using the following relation:

\begin{equation}
\mathrm{r_B~(kpc) = 0.031~\frac{M_{BH}~(M_\odot)}{kT~(keV)}}
\end{equation}
\\
\noindent
using kT = 1 keV \citep[e.g.,][]{Russell2015}. The calculated Bondi radius $\sim20$~parsec was approximately the jet bending radius estimated by \citet{Bondi2004} for 1ES~2344+514. This could indicate that the jet flaring occurs as it is leaving the Bondi radius, and the jet bending indicates the presence of dense medium around 20 parsec and beyond. Overall, this is consistent with the idea in the literature that BL Lac jets may be undergoing stronger jet-medium interactions that may destabilise their jets, compared to quasars \citep{Urry1995}.

\subsubsection{The Nature of the `Eastern' Source}
A radio source is located around 160 kpc to the east of the radio core in 1ES~2344+514. This source has previously been observed in VLA observations \citep[e.g.,][]{Giroletti2004}. This source is clearly resolved into two sub-components in the uGMRT image. The nature of this source has been unclear and there have been suggestions that it could be associated with 1ES~2344+514 as extended low-brightness emission is present between the core and this eastern source \citep{Giroletti2004}. Our image shows that it has a high fractional polarization and the EVPA vectors are consistent with the two sub-components being AGN hotspots: the perpendicular B-fields could be indicating B-field compression in terminal shocks (Figure~\ref{fig:1ES2344z}). The discovery of X-ray emission coincident with the eastern source \citep{Donato2003} supports the picture of this source being a background double-lobed radio galaxy. 

\subsection{Quasar 4C71.07}
\begin{figure}
  \centering
  \includegraphics[width=8.6cm]{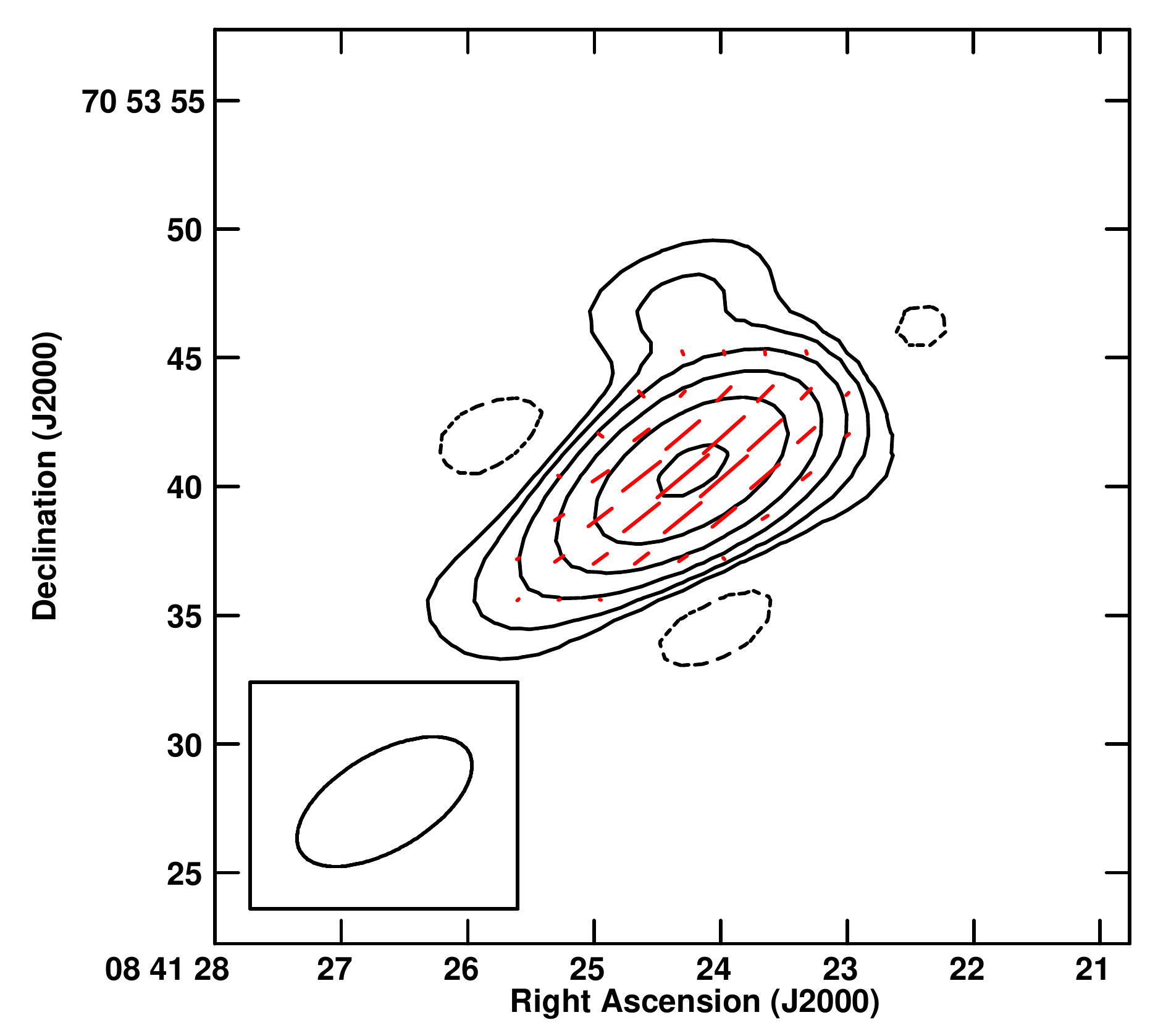}
    \includegraphics[width=8.3cm]{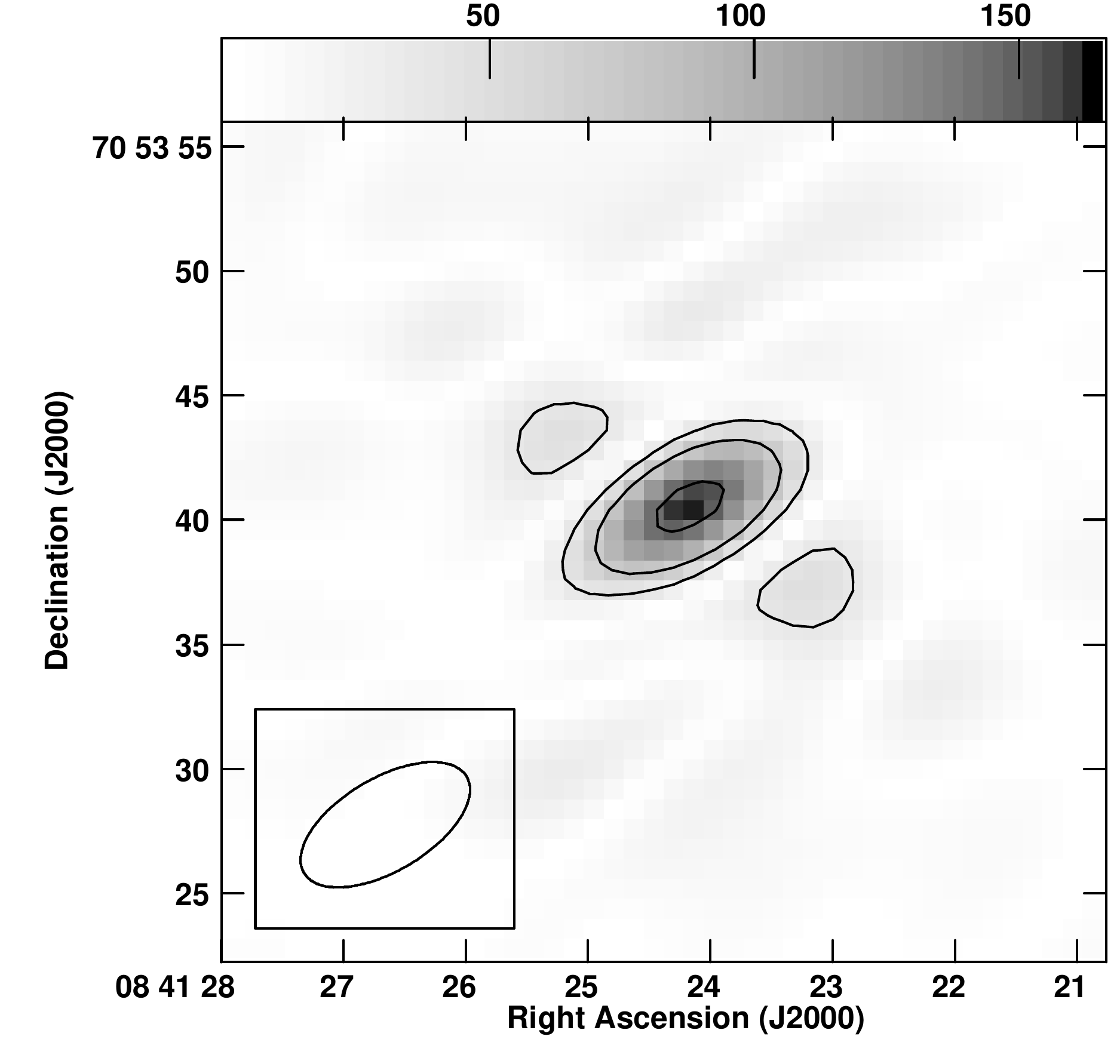}
  \caption{{  (Left) 612 MHz contour image of the quasar 4C71.07 with EVPA vectors. The beam is $7.47'' \times 3.78''$ at a PA of $-59^\circ$. The peak surface brightness ($I_P$) is $4.07~$Jy~beam$^{-1}$. The contour levels in percentage of the peak surface brightness $I_P$ are $(\pm2.8,~5.6,~11.25,~22.5,~45,~90)$~Jy~beam$^{-1}$. The length of the EVPA vectors is proportional to polarized intensity with $5''$ corresponding to 312~mJy~beam$^{-1}$. (Right) 612 MHz polarized intensity in contours and greyscale of the quasar 4C71.07 with units in mJy~beam$^{-1}$. The  peak polarised surface brightness ($P_P$) is 0.16~Jy~beam$^{-1}$. The contour levels in percentage of the peak polarised surface brightness $P_P$ are $(22.5,~45,~90)$~Jy~beam$^{-1}$.}}
  \label{fig:4C71.07}
\end{figure}

A radio-optical overlay of the quasar 4C71.07 has been presented in Figure \ref{fig:radio_optical}; the DSS optical image in colour is superimposed by uGMRT 612 MHz contours. The 612 MHz image of 4C71.07 is presented in Figure \ref{fig:4C71.07}. The total intensity radio contours are superimposed by the EVPA vectors as red ticks. The synthesised beam is $7.47''\times 3.78''$ at a PA of $-59^\circ$. The rms noise in the total intensity image is { 38 mJy~beam$^{-1}$ } and polarized intensity image is { 7.1}~mJy~beam$^{-1}$. No extended emission is detected in the uGMRT image of 4C71.07. The core however is linearly polarized. The 4.86~GHz VLA image of 4C71.07 by \citet{ODea1988} shows a core and a south-western hotspot. Our uGMRT image does not resolve the core-hotspot structure but shows linear polarization consistent with the image of \citet{ODea1988}, with the core dominating the polarized emission. Assuming optically thick emission in the core, we find that the inferred B-field is transverse to the jet direction. { Interestingly, signatures of a helical B-field on parsec-scales derived via RM gradients have been noted for 4C71.07 by \citet{Asada10}.}

\subsection{Minimum Energy B Field}
We estimated the `minimum energy' magnetic field ($B_{min}$) in the two blazars showing extended (lobe) emission, viz., 3C390.3 and 1ES~2344+514, using the relations in \citet{O'Dea1987}:

\begin{equation}
    B_{min} = [2\pi(1+k)C_{12}L_{rad}(V\phi)^{-1}]^{2/7}~~~\mathrm{G}
\end{equation}
where
\begin{equation}
L_{rad} = 1.2 \times 10^{27}  D_{Mpc}^2 S_0 \nu_0^{-\alpha}(1+z)^{-(1+\alpha)}
     \times (\nu_u^{1+\alpha} - \nu_l^{1+\alpha})(1+\alpha)^{-1}~~~\mathrm{ergs ~s^{-1}}
\end{equation}
\\
\noindent
and $k$ is the ratio of relativistic proton to relativistic electron energy, $V$ is the source volume, $C_{12}$ is a constant which depends on the spectral index and frequency cutoffs \citep[here, $C_{12} = 2.0 \times 10^7$;][]{Pacholczyk1970}, $L_{rad}$ is the total radio luminosity, $\phi$ is the volume filling factor, $z$ is the source redshift, $D_{Mpc}$ is the luminosity distance of the source in Mpc, $S_0$ is the flux density in Jy at a fiducial frequency $\nu_0$ in Hz, $\alpha$ is the spectral index, and $\nu_u$ and $\nu_l$ are the upper and lower frequency cutoffs in Hz. 
Using the following assumptions:
\begin{enumerate}
    \item the radio spectrum extends from 100 MHz to 15 GHz with a spectral index of $\alpha=-0.65$.
    \item the relativistic protons and relativistic electrons have
equal energies (k = 1).
    \item the jets are cylinders uniformly filled with relativistic particles and magnetic fields ($\phi$ = 1)
\end{enumerate}
the minimum energy B field was calculated for 3C390.3 and 1ES~2344+514 (see Table \ref{table2}). These estimates are similar to those obtained in other radio galaxies and blazars \citep[e.g.,][]{Stawarz2005,OSullivan2009}.

\begin{table}
\begin{center}
\tabularfont
\caption{Minimum energy B field ($B_{min}$) estimates for the extended emission in 2 blazars}\label{table2} 
\begin{tabular}{lclcccc}
\topline
Source & Region & & Extent~(kpc)& $B_{min}~(\mu G)$ & \\\midline
3C390.3 & Lobes : & North-west & { 78} & { 1.94}\\
&& South-east & { 90} & { 1.92}\\ \hline
1ES~2344+514 & Diffuse Halo : & &96 & 1.20 \\\hline
\end{tabular}
\end{center}
\end{table}

\section{Summary and conclusions}
We have carried out Band 4 ($\sim600$~MHz) uGMRT polarization observations of 3 blazars and detected extensive linear polarization in all of them. Below we summarise our primary results from this study.

\begin{enumerate}
\item The quasar 3C390.3 exhibits extensive polarization in its core, lobes and hotspots. The BL Lac object 1ES~2344+514 and the quasar 4C71.07 exhibit polarized emission primarily associated with their radio cores.

%\item The degree of linear polarization is similar in the cores of the two quasars and the BL Lac object. This implies that any depolarizing medium is confined to the 100 pc to kpc scales in these blazars since VLBI observations do indicate clear differences in the nuclear environments of quasars versus BL Lac objects.

\item {  The degree of linear polarization in the cores of the two quasars is higher than in the BL Lac object in our study. This is consistent with differences in the nuclear environments of quasars and BL Lac objects, as previously indicated in VLBI observations of blazars in the literature.}

\item A rotation of EVPA vectors is observed in the northern hotspot region of the quasar 3C390.3. This could indicate jet precession that could arise due to accretion disk instabilities or the presence of binary supermassive black holes.

\item Greater depolarization in the south-eastern lobe and hotspot of the quasar 3C390.3 compared to the northern one indicates that the north-western jet is the approaching one, and the source is undergoing the `Laing-Garrington effect'. 

\item The core polarization structure in the BL Lac 1ES~2344+514 is complex but could be consistent with a jet bend smaller than the uGMRT beam. Jet bending could suggest a much denser environment around the BL~Lac 1ES~2344+514 on the 100 parsec- to the kpc-scale. 
%This is consistent with the suggestion of denser local environments in BL Lacs compared to quasars in the literature. 
Stronger jet medium interaction may result in the disruption of jets in BL Lac objects.

\item Using polarization data, we are able to rule out a compact source to the east of 1ES~2344+514 as being associated with 1ES~2344+514. Instead, its polarization structure suggests it to be a background double radio galaxy with polarized hotspots.

\item For 4C71.07, we see the core polarization dominating the overall emission in this unresolved source with inferred B-fields being transverse to the jet direction as seen in the literature.
 \end{enumerate}

Multi-frequency uGMRT polarimetric data are needed to study the kpc-scale rotation measures across these blazars in order to look for differences in their surrounding media. Such studies are currently underway and will be presented in our upcoming papers. 

\section*{Acknowledgements}
We thank the staff of the Giant Metrewave Radio Telescope (GMRT) who have made these observations possible. The GMRT is run by the National Centre for Radio Astrophysics of the Tata Institute of Fundamental Research. We acknowledge the support of the Department of Atomic Energy, Government of India, under the project 12-R\&D-TFR-5.02-0700.

\typeout{}
\bibliography{latest_JAA_paper}
\end{document}